\documentclass[a4paper]{article}

\usepackage[english]{babel}
\usepackage[utf8x]{inputenc}
\usepackage{amsmath}
\usepackage{graphicx}
\usepackage[colorinlistoftodos]{todonotes}

\title{How many bits specify a quantum state?}
\author{Barbara Drossel, Institute of Condensed Matter Physics, TU Darmstadt}

\begin{document}
\maketitle

\hskip 1cm \parbox{9cm}{\small  Quantum mechanics suggests that nature is discrete, with one state per phase space volume $\hbar^{3N}$. This appears to contradict the idea that the state of an $N$-particle system can have infinite precision and is described by a set of exponentially many complex numbers. Using a finite-temperature gas confined in a box as an example, this short paper argues that there are indeed limits to the precision of wave functions, and that this may help at understanding the quantum-to-classical transition.} 

\medskip

\medskip

\medskip

Imagine a gas of $N\simeq 10^{23}$ atoms in a rectangular box. A hundred years ago, the microscopic behavior of such a gas was described by classical mechanics, with atoms being represented as  small balls with a short-range interaction. The time evolution of this set of balls is described by a trajectory in $6N$-dimensional phase space. Since the dynamics is highly chaotic, prediction of the time evolution beyond some short time horizon is impossible, and therefore probabilistic considerations were applied, supposing that all possible trajectories that start within the same small phase space volume are equally probable. However, such probabilistic arguments are foreign to a deterministic theory, and famous discussions relating to classical statistical mechanics were triggered by this conceptual problem\cite{steckline1983zermelo}. Indeed, from a physics point of view an infinite-precision random initial state that is combined with a deterministic time evolution cannot be distinguished from a finite-precision initial state that is combined with a time evolution that chooses at random the bits that are not determined by the initial state while the trajectory develops\cite{gisin2016}. 

The advent of quantum mechanics gave support to this latter view, as it refutes the idea of infinitely precise phase-space points. The uncertainty principle states that points in phase space have a finite size of the order $\hbar^{3N}$. Combined with the discovery of true randomness on the quantum level, this suggests that the time evolution of the atomic gas is not merely random from the perspective of an observer who only has a limited knowledge of the initial state, but that this randomness is an inherent feature of the dynamics, with the future not being determined by the present state beyond the time horizon associated with an initial precision $\hbar^{3N}$. 

However, the monsters of full determinism and infinite precision were only temporarily killed by these insights. They rose again in different disguise and became much worse than in classical mechanics. Instead of phase-space trajectories, now the state vector, in combination with the Schr\"o\-dinger equation is widely believed to fully specify the future time evolution of a nonrelativistic many-particle system. Evidence for the reality of the wave function\cite{pusey2012reality} and its wide scope\cite{nairz2003quantum,herbst2015teleportation} is accumulating, weakening interpretations of quantum mechanics that view the state vector merely as a representation of our state of knowledge or of an ensemble of systems, or as applying only to the microscopic scale. The wave function can be measured with tomographic methods\cite{vallone2016} and can even be precisely shaped using feedback mechanisms\cite{weinacht1999}. Due to these successes, many quantum physicists take the view that a quantum state, combined with unitary time evolution describes also systems that consist of a macroscopic number of particles\cite{gogolin2016equilibration}, and possibly even the entire universe\cite{hartle1983wave}. 

The conclusions and paradoxes resulting from this view are extremely problematic \cite{adler2003decoherence,schlosshauer2005decoherence,ellis2012}. Using as a basis the products of $M$ possible one-particle states, the specification of a state of a $N$-particle system requires a set of $M^{N}$ complex numbers, which is far more than the $6N$ real numbers specifying a phase space point in classical mechanics. 
It implies an enormous amount of entanglement between all particles that have ever interacted in the past.  This huge entanglement cannot be measured and is in contradiction with the emergence of classical objects, where each atom has its pretty well specified position and appears to be free from entanglements and superpositions. It also appears at odds with the original insight of quantum physics that nature is specified by a limited number of bits. 

Let us therefore critically examine the claim that the wave function has infinite precision. There are in fact only few conditions under which a quantum state is well defined from an empirical point of view\cite{ellis2012}. First, the wave function of a $N$-particle system is well defined when the system is in the ground state or in a long-lived eigenstate of a Hamiltonian. This state can be specified by the label $n$ that indicates the eigenstate. Since the density of eigenstates can be approximately obtained from Bohr's quantization rule, this means that the number of bits to specify the state given the Hamiltonian of the system is of the order of the number of bits required to fix a cell of size $\hbar^{3N}$ in phase space (plus those required to fix the spin). Second, a wave function is well defined when the procedure of preparing it is well specified and repeatable. In this case, no further bits are required to characterize the state.  However, in the vast majority of cases a nonzero temperature prevents the full controllability of the wave function, and this is the basis of the research fields of decoherence and open quantum systems. 

In order to discuss the nature of thermalized states, let us come back to the simplest finite-temperature system, the gas of $10^{23}$ atoms in a box. This gas is not in an eigenstate of the Hamiltonian, as its energy (including the uncertainties) covers a huge amount of energy levels. Neither can the initial state of this gas be controlled. Even when each atom is prepared carefully by some emission procedure that sends it into the box, the emitter must be measured (for instance by measuring its recoil) in order to make sure that the atom has been emitted. This measurement involves a thermal environment and introduces therefore some uncertainty in the precise shape of the wave function of the emitted atom. Of course many physicists would argue that this uncertainty is only due to our limits of knowledge. However there is no way even in principle to eliminate this uncertainty by better observation. 

Let us therefore explore the logical consequences of accepting an objective uncertainty of the wave function of the gas (in addition to subjective uncertainties due to a lack of information, which we do not discuss here). Any initial uncertainty  will increase under time evolution in this non-integrable system, implying that the initial state, together with the Hamiltonian, cannot fix the future states beyond some time horizon. This does not mean that a future state has no specific microscopic features at all, but that these features are not specified by the initial state. In fact, the intuition underlying the Boltzmann formula for entropy $S=k_B \ln \Omega$ is that the state of an equilibrated gas is one out of a number of $\exp(S/k_B)$ possible states through which the system moves by a stochastic process. Furthermore, quantum statistical mechanics tells us that this number of states is identical to the above-mentioned number of phase space volume cells of size $\hbar^{3N}$. If we take this at face value, the gas evolves by ongoing nonunitary, stochastic localization to one of the possible configurations. But statistical mechanics in itself gives no indication of the nature of these configurations, since the density matrix of an isolated system is proportional to the identity matrix in all bases. Nevertheless, we have strong evidence that the finite-temperature gas is appropriately described by localized wave packets of a spatial size of the order of the thermal wave length $\lambda_{th}=h\sqrt{2\pi m k_BT}$\cite{tenreasons}: First, decoherence theory states that the preferred basis of a particle embedded in a thermalized environment is the position basis, since the interaction potential is position dependent\cite{schlosshauer2005decoherence}. Second, molecular dynamics simulations, which assume well localized atoms, are extremely successful at describing systems that consist of many atoms at finite temperature\cite{marx2000ab}. Third, as temperature is decreased, a phase transition to a solid occurs, where the particles become definitely localized. Fourth, the requirement that the transition to classical mechanics shall be continuous suggest that the particles become increasingly localized with increasing temperature. 

The picture that emerges from these considerations is as follows: Many-particle states that are far away from the ground state cannot be controlled or precisely determined even in principle and thus have only a limited precision. Finite precision implies stochasticity, as only one of the many future states compatible with an initial state becomes realized. It also implies a limited amount of entanglement, as a huge amount of entanglement requires a correspondingly larger number of bits for its description.  Each particle of such a system is localized in a small region of space, a situation that is similar to the one described by  continuous collapse theories\cite{bassi2013}. If the system is in thermal equilibrium, the number of bits that specify the state is proportional to the entropy, and localization is within a distance of the order of the thermal wave length, which thus also sets a cutoff for entanglement. 

If additionally the particle density of the gas is much lower than $1/\lambda_{th}^3$,  quantum statistical effects in the gas can be neglected\cite{kittel1980thermal}. In this case the classical and quantum mechanical descriptions of the gas converge, because both include localized particles and stochasticity if we accept a finite precision. This represents a good starting point for understanding the quantum-classical transition, because particle localization as well as stochastic choices of one option over another are required for the formation of macroscopic objects that consist of a specific arrangement of atoms. And, finally, they are also required for the measurement process.

\medskip
{\small I thank Steven Weinstein, Lee Smolin, and George Ellis for useful comments on the manuscript, and the Perimeter Institute for hosting me during my sabbatical.  Perimeter Institute is supported by the Government of Canada  through the Department of Innovation, Science and Economic Development and by the Province of Ontario through the Ministry of Research, Innovation and Science.}

\end{document}